%% file: Maedaetal_2022_CPD2-orbit_En.tex
\pdfoutput=1
\documentclass[preprint]{aastex63}
\usepackage{mathtools}
\usepackage{color}
\usepackage{multirow}
\usepackage{comment}

\turnoffeditone

\usepackage{bm}
\usepackage{here}
\usepackage{textcomp} 
\usepackage{physics}
\usepackage{comment}
\usepackage{mathtools}

\newcommand{\mr}{\mathrm}
\renewcommand{\thefootnote}{*\arabic{footnote}}

\newcommand{\Equref}[1]{Equation~(\ref{#1})}
\newcommand{\Figref}[1]{Figure~\ref{#1}}
\newcommand{\Secref}[1]{\S~\ref{#1}}

\newcommand{\Equrefs}[1]{Equations~(\ref{#1})}

\received{December 18, 2023}
\accepted{April 16, 2024}
\submitjournal{Astrophysical Journal}

\shorttitle{Delivery of Dust particles into Circumplanetary Disk}
\shortauthors{Maeda et al.}

\begin{document}

\title{Delivery of Dust Particles from Protoplanetary Disks onto Circumplanetary Disks of Giant Planets}

\correspondingauthor{Natsuho Maeda}
\email{nmaeda@tiger.kobe-u.ac.jp}

\author[0000-0002-6874-5178]{Natsuho Maeda}
\affiliation{Department of Planetology, Kobe University, 1-1 Rokkodai-cho, Nada-ku, Kobe 657-8501, Japan}

\author[0000-0002-4383-8247]{Keiji Ohtsuki}
\affiliation{Department of Planetology, Kobe University, 1-1 Rokkodai-cho, Nada-ku, Kobe 657-8501, Japan}

\author[0000-0002-8478-1881]{Ryo Suetsugu}
\affiliation{National Institute of Technology, Oshima College, Oshima 742-2193, Japan}

\author[0000-0003-2993-5312]{Yuhito Shibaike}
\affiliation{Space Research and Planetary Sciences, Physics Institute, University of Bern, CH-3012 Bern, Switzerland}
\affiliation{National Astronomical Observatory of Japan, 2-21-1 Osawa, Mitaka, Tokyo 181-8588, Japan}

\author[0000-0002-5964-1975]{Takayuki Tanigawa}
\affiliation{National Institute of Technology, Ichinoseki College,
Ichinoseki 021-8511, Japan}

\author[0000-0002-0963-0872]{Masahiro N. Machida}
\affiliation{Department of Earth and Planetary Science, Graduate School of Science, Kyushu University, 
Fukuoka 819-0395, Japan}

\begin{abstract}
Principal regular satellites of gas giants are thought to be formed by the accumulation of solid materials in circumplanetary disks (CPDs). While there has been significant progress in the study of satellite formation in CPDs, details of the supply of satellite building blocks to CPDs remain unclear. We performed orbital integration of solid particles in the protoplanetary disk (PPD) approaching a planet, considering the gas drag force using the results of three-dimensional hydrodynamical simulations of a local region around the planet.
We investigated planetary-mass dependence of the capture positions and capture rates of dust particles accreting onto the CPD. 
We also examined the degree of dust retention in accreting gas onto the CPD, which is important for determining the ratio of dust-to-gas inflow rates, a key parameter in satellite formation.
We found that the degree of dust retention increases with increasing planetary mass for a given dust scale height in the PPD. In the case of a small planet ($M_{\rm p}=0.2M_{\rm Jup}$), most particles with insufficient initial altitudes in the PPD are isolated from the gas in the accreting region. On the other hand, in the case of a massive planet ($M_{\rm p}=1M_{\rm Jup}$), dust particles can be coupled to the vertically accreting gas, even when the dust scale height is about $10-30$\% of the gas scale height. 
The results of this study can be used for models of dust delivery and satellite formation in the CPDs of gas giants of various masses, including exoplanets.
\end{abstract}

\section{INTRODUCTION} \label{sec:intro}
\input{01introduction}

\section{METHODS}\label{sec:methods}
\input{02method}

\section{RESULTS}\label{sec:results}
\input{03result}

\newpage
\section{DISCUSSION}\label{sec:discussions}
\input{04discussion}

\section{CONCLUSIONS}\label{sec:conclusions}
\input{05conclusion}

\vspace{1cm}
We thank Yuri Fujii and Hidekazu Tanaka for helpful comments and discussion. We also thank the anonymous referee for helpful comments. This work was supported by JSPS KAKENHI Grant numbers JP22J10202, JP18K11334, JP21H00043, JP22H01286, JP19K14787, JP15H02065, JP20K04051, JP22H01274.
N.M. gives thanks for the support by the Kobe University Doctoral Student Fellowship and JSPS Research Fellowship for Young Scientists.
Y.S. contributed to this work within the framework of the NCCR PlanetS supported by the Swiss National Science Foundation under grants 51NF40\_182901 and 51NF40\_205606, and acknowledges support from the Swiss National Science Foundation under grant 200021\_204847 `PlanetsIn-Time'.
This research used the computational resources of the 2020, 2021, 2022 and 2023 Koubo Kadai on Earth Simulator (NEC SX-ACE) at JAMSTEC, supercomputing resources at Cyberscience Center, Tohoku University, and the general-purpose PC cluster at the Center for Computational Astrophysics, National Astronomical Observatory of Japan.



\input{bib}
\end{document}

%% file: 01introduction.tex
Massive regular satellites, such as the Galilean satellites and Titan, are thought to be formed by accumulation of solids in circumplanetary disks (CPDs), gaseous disks formed around giant planets when the planet accretes gas of the protoplanetary disks (PPDs).
In early models of satellite formation around Jupiter and Saturn, their CPDs were assumed to be isolated from the PPD. The total dust mass in each CPD is assumed to be consistent with the total current mass of the satellites in each system, which is a similar concept to the minimum mass solar nebula (MMSN) model \citep{ls82}.
\citet{me03a,me03b} extended this model by considering CPDs composed of an optically 
thick inner region and an optically thin outer region.
Satellite formation in such CPDs has been recently examined in 
detail by \citet{mi16}. They found that the high gas density in the inner disk leads to very rapid migration of satellitesimals, and it is difficult to form the current Galilean satellites in such a CPD.

On the other hand, \citet{cw02} proposed the gas-starved disk model for the CPD, in which gas and solids are continuously supplied from the PPD.
\citet{cw06} showed that the gas-starved disk can reproduce the common mass ratio of the satellite system to the host planet observed for giant planets in the solar system.
In addition, the hydrodynamic simulations showed that a CPD is naturally formed around a sufficiently massive planet during its gas accretion phase \citep[e.g.,][]{tw02,m08,t12,sz14,sz16,sz17}. Thus the picture of the gas-starved disk has been widely accepted, and various studies of satellite formation have been conducted with such CPDs \citep[e.g.,][]{cw02,cw06,s10,oi12}.
Recently, a model of satellite formation in a decretion disk, which has nearly vertical inflow in the inner part and outflow near the midplane, has also been proposed \citep{bm20}.

In such models of satellite formation, the total amount and the size of solid particles supplied to the CPD are important, because they likely determine the timescale of satellite accretion and the composition of the formed satellites \citep[e.g.,][]{cw02,r17}.
\citet{f13} studied capture of kilometer-size and larger planetesimals, and \citet{so17} examined their capture and subsequent orbital evolution in the CPD. Such planetesimals can be captured in the inner part of the CPD where the gas density is sufficiently high and the planetesimals suffer strong gas drag. \citet{so17} concluded that captured planetesimals would contribute to satellite formation more than the dust supplied with the gas if both the velocity dispersion of the planetesimals and the width of the planetesimal gap in the PPD are sufficiently small.
If planetesimals can be captured near the outer edge of the gas gap due to the filtering effect \citep[e.g.,][]{r06}, external gravitational perturbation from other planets or planetary embryos would be necessary to deliver them to the CPD \citep{r18}.
On the other hand, relatively small particles such as pebbles and dust would also be important building blocks of satellites.
\citet{sh19} proposed the so-called slow-pebble-accretion scenario of Galilean satellite formation, where a small number of planetesimals captured by the CPD slowly accretes pebbles supplied into the CPD. The scenario can reproduce most characteristics of the current Galilean satellites if some reasonable parameters are assumed.
While many of these studies assumed sufficient solids in the CPD, there are still many unknowns about the supply of solid particles into the CPD.

Since meter-sized and smaller particles are much more susceptible to the effects of gas drag than kilometer-size planetesimals, the gas field around the CPD needs to be taken into account for their delivery into the CPD.
High-resolution hydrodynamic simulations showed that only the off-midplane gas in the PPD can accrete onto the CPD and the accretion occurs nearly vertically. In addition, part of the accreted gas flows radially outward through the planet's Hill sphere on the midplane unless the viscosity of the PPD is very small \citep[e.g.,][]{t12,sz14,sz17}, which prevents small particles confined in the midplane of the PPD from accreting into the CPD \citep{t14}.
Thus, assistance of vertical inflow is important for the delivery of small particles coupled with the gas to the CPD.
\citet{h20} investigated the supply to the CPD of solid particles with various sizes distributed vertically in the PPD, considering the gas field around the planet obtained from the local hydrodynamic simulation of \citet{t12}. They found that small particles sufficiently stirred up by turbulence in the PPD can accrete onto the CPD with the vertically accreting gas.


However, \citet{t12,t14} and \citet{h20} only examined the case where the Hill radius of the planet is the same as the scale height of the PPD, which corresponds to a planet with the mass of $0.4M_{\rm Jup}$ ($M_{\rm Jup}$ is the current mass of Jupiter) at 5.2~au in the case of the MMSN, and the planetary-mass dependence was not investigated.
Understanding of the planetary-mass dependence is important for the understanding of the satellite formation around planets with various masses, for example, Jupiter, Saturn, and exoplanets.
Also, CPDs of exoplanets have been observed recently \citep[e.g.,][]{i19}, and a better understanding of the planetary mass dependence is expected to contribute to modeling of solid delivery into the CPD for comparison with the observations.

In our recent paper \citep{m22}, we performed local three-dimensional high-resolution hydrodynamic simulations for various planetary masses (corresponding to $0.05-1M_{\rm Jup}$ at 5.2~au for the MMSN) with the same numerical code used in \citet{t12} \citep[see also][]{m08}, and revealed the planetary-mass dependence of the source region and mass accretion rate of gas accreting into the CPD.
In the present work, we will examine the planetary-mass dependence of the supply of solid particles into the CPD using the gas fields obtained by \citet{m22}.

The strucuture of this paper is as follows. In \Secref{sec:methods}, we will explain basic equations, gas drag calculation, and numerical settings. In \Secref{sec:results}, we will show the results of orbital calculation, such as initial altitudes and radial locations in the PPD of dust particles captured in the CPD, and capture rates of dust particles.
In \Secref{sec:discussions}, we will derive the capture rate as a function of dust scale height, and also examine the degree of dust retention in accreting gas into the CPD. 
We will also derive the dust mass accretion rate onto the CPD and the ratio of dust-to-gas inflow rates by combining our results with the gap models for the surface densities of dust and gas obtained in previous works.
We will also discuss model limitations of this study.
We will summarize this work in \Secref{sec:conclusions}.

%% file: 02method.tex
\subsection{Basic Equations}
We integrate orbits of particles taking account of the effects of the gas flow perturbed by the gravity of a planet, using the methods described in \citet{h20} \citep[see also][]{t14}.
We consider a three-body problem for a central star, a planet, and a particle, assuming that the planet is on a circular orbit.
We adopt the rotating local Cartesian coordinate system with origin at the planet, the $x$-axis pointing radially outward, the $y$-axis in the direction of the planet's orbital motion, and the $z$-axis in the vertical direction to the PPD midplane. We neglect the curvature of the protoplanetary disk, and define $r$ and $R$ as $r\equiv (x^2+y^2+z^2)^{1/2}$ and $R\equiv (x^2+y^2)^{1/2}$, respectively. 
When the masses of the planet and the particle are sufficiently smaller than that of the central star and the orbital eccentricity and inclination are small, the motion of the particle relative to the planet can be described by the Hill equations \citep{n89,o12}.
The Hill equations can be normalized by the planet's Hill radius $r_{\mr{H}}=(M_{\mr{p}}/(3M_{\mr{c}}))^{1/3}a$ and the inverse of orbital angular velocity of the planet $\Omega_{\mr{K}}^{-1}=(GM_{\rm c}/a^3)^{-1/2}$, where $M_{\rm c}$, $M_{\rm p}$, and $a$ are the masses of the central star and the planet, and the semi-major axis of the planet, respectively.
The normalized Hill equations considering gas drag are described as \citep{f13,t14,h20}:
\begin{equation}\label{eq:orbitEOM}
\begin{dcases}
\ddot{\tilde{x}}= \;\;\; 2\dot{\tilde{y}} +3\tilde{x}-\frac{3\tilde{x}}{\tilde{r}^3} +\tilde{a}_{\mr{drag},x}, \\
\ddot{\tilde{y}}=-2\dot{\tilde{x}} \qquad \; -\frac{3\tilde{y}}{\tilde{r}^3} +\tilde{a}_{\mr{drag},y}, \\
\ddot{\tilde{z}}= \qquad \quad -\tilde{z}-\frac{3\tilde{z}}{\tilde{r}^3} +\tilde{a}_{\mr{drag},z},
\end{dcases}
\end{equation}
where tildes denote normalized quantities.

Using the gas drag force $\bm{F}_{\rm drag}=-(1/2)C_{\rm D}\pi r_{\rm d}^2 \rho_{\rm g} \Delta u \Delta \bm{u}$ and the dust particle mass $m_{\rm d}$, the acceleration due to the gas drag force $\bm{a}_{\mr{drag}}= \bm{F}_{\rm drag}/m_{\rm d}$ is expressed in the normalized notation as
\begin{eqnarray}
\tilde{\bm{a}}_{\mr{drag}}=
-\frac{3}{8}C_{\mr{D}}\frac{\rho_{\mr{g}}}{\rho_{\mr{d}}}\tilde{r}_{\mr{d}}^{-1}\Delta \tilde{u}\Delta \tilde{\bm{u}},
\end{eqnarray}
where $C_{\mr{D}}$, $\rho_{\mr{d}}$, $\rho_{\mr{g}}$, $r_{\mr{d}}$, $\Delta \bm{u}$ are the gas drag coefficient, bulk density of the particle (we assume $\rho_{\rm d}=10^3$~$\rm{kg \; m^{-3}}$), gas density, particle radius, and the velocity of the particle relative to the gas, respectively.

\subsection{Background Gas Flow}
\subsubsection{Hydrodynamic Simulations}
We use the gas fields obtained by \citet{m22} with local three-dimensional hydrodynamical simulations to calculate the gas drag force for dust particles.
The nested-grid method \citep[e.g.,][]{m05,m06} is used for the hydrodynamical simulations to examine the gas flow in the vicinity of the CPD with high resolution while considering a wide area around the planet ($x,y\in[-12h_{\mr{g}},\; 12h_{\mr{g}}]$, $z\in[0,\; 6h_{\mr{g}}]$, where $h_{\rm g}$ is the gas scale height of the PPD).
We denote the level of the nested-grid by $l$, where $l=1$ corresponds to the whole simulation box. With each increment of the level of the nested-grid, the size of the computational box is reduced by a factor of two, keeping the center of the box and the number of cells the same; the number of the cells for all grid levels is $(n_x,n_y,n_z)=(64,64,16)$. We set the maximum grid level to be 11.

An important parameter in our simulation is the Hill radius of the planet normalized by the scale height of the PPD, denoted by
\begin{equation}
\frac{r_{\mr{H}}}{h_{\mr{g}}}=\left( \frac{M_{\rm p}}{3M_{\rm c}} \right)^{1/3} \frac{a}{h_{\rm g}}.
\end{equation}
Assuming that $M_{\rm c}=M_{\odot}$ and the temperature distribution of the PPD is given by the MMSN model \citep{h81} as
\begin{equation}\label{eq:T}
T=280\; \mr{K}\;(a/1~\mr{au})^{-1/2},
\end{equation}
we obtain the relationship between the planetary mass $M_{\rm p}$ and the parameter $r_{\mr{H}}/h_{\mr{g}}$ as
\begin{equation}
\frac{M_{\rm p}}{M_{\rm Jup}}=0.41\left( \frac{M_{\rm c}}{M_{\odot}} \right)^{1/2} \left( \frac{a}{5.2~\mr{au}} \right)^{3/4} \left(\frac{r_{\rm H}}{h_{\rm g}}\right)^{3}.
\end{equation}
Thus, $r_{\mr{H}}/h_{\mr{g}}=0.8$, 1.0, and 1.36 correspond to the planetary masses of $M_{\mr{p}}=0.2$, 0.4, and 1.0$M_{\mr{Jup}}$, respectively.
In this work, we will examine these three cases.

\subsubsection{Effects of Gas Drags}
As for the gas drag coefficient $C_{\mr{D}}$, we use the following interpolation formula that reproduces known expressions in limiting cases \citep[see][]{t14,h20}:
\begin{equation}
C_{\mr{D}} = \left[ \left(\frac{24}{\mathcal{R}} + \frac{40}{10+\mathcal{R}} \right)^{-1} + 0.23\mathcal{M} \right]^{-1} + \frac{(2.0-w_{\mr{c}})\mathcal{M}}{1.6+\mathcal{M}} + w_{\mr{c}},
\end{equation}
where $\mathcal{R}=2r_{\mr{d}}\Delta u/\nu$ is the Reynolds number, and $\mathcal{M}=\Delta{u}/c_{\mr{s}}$ is the Mach number. A correction factor $w_{\mr{c}}$ is defined so that $w_{\mr{c}}=0.4$ for $\mathcal{R}<2\times 10^5$ and $w_{\mr{c}}=0.2$ for $\mathcal{R}>2\times 10^5$. The kinetic viscosity is given by $\nu=0.353\sqrt{8/\pi}c_{\mr{s}}l_{\mr{g}}$ \citep{cc70}, where $l_{\mr{g}}=m_{\mr{mol}}/(\sigma_{\mr{mol}}\rho_{\mr{g}})$ is the mean free path.
We set the mass and collision cross section of a gas molecule to $m_{\mr{mol}}=3.9\times 10^{-27} \; \mr{kg}$ and $\sigma_{\mr{mol}}=2.0 \times 10^{-19}\; \mr{m^2}$, respectively.

In the orbit calculation, we obtain the gas density $\rho_{\mr{g}}$ and the relative velocity between gas and dust particles $\Delta\bm{u}$ at the position of the particle from the results of hydrodynamic simulation, and calculate the gas drag force at each time step.
We assume the gas surface density at Jupiter's orbit in the MMSN model \citep{h81},  $\Sigma_{\rm g,0}=1434\; \mr{kg\; m^{-2}}$, for the unperturbed region of the PPD.

While approaching the planet, sufficiently large particles undergo vertical oscillation across the midplane of the PPD due to the vertical component of the tidal force.
In the case of the Epstein drag law, the critical radius $r_{\mr{d,crit}}$ for this vertical oscillation can be expressed as $r_{\mr{d,crit}}=v_{\mr{th}}\rho_{\mr{g}}/(2\rho_{\mr{d}}\Omega_{\mr{K}})$, where $v_{\mr{th}}=\sqrt{8/\pi}c_{\mr{s}}$ is the thermal velocity of the gas.
If we assume that the planet is located at 5.2~au from the central star in the MMSN model and the depletion factor of the gas surface density is given by $f_{\rm H}$ ($f_{\rm H}=1$ corresponds to the MMSN model), $v_{\mr{th}}=1.1\times 10^{3}\; \mr{m\; s^{-1}}$, and $\Omega_{\mr{K}}=1.7\times 10^{-8} \; \mr{rad\; s^{-1}}$, and the gas density at the midplane is $\rho_{\mr{g}}=1.5\times 10^{-8} f_{\mr{H}}$ $\mr{kg\; m^{-3}}$, thus we obtain $r_{\mr{d,crit}}\simeq 0.5f_{\mr{H}}$~m.
For example, when a gap is opened around the planetary orbit and $f_{\mr{H}}=0.1$, we have $r_{\mr{d,crit}}\simeq 5$~cm.
We examined cases with $r_{\rm d}=10^{-4}-10$~m, but we will mostly focus on the cases of $r_{\mr{d}}=10^{-4}-10^{-2}~\mr{m}$ ($< r_{\mr{d,crit}}$ when $f_{\rm H}\gtrsim 0.02$), where the particles are easily coupled to the gas accreting into the CPD.

\subsection{Settings of Orbital Integration for Solid Particles}
We integrate \Equref{eq:orbitEOM} using the eighth-order Runge-Kutta integrator as used in \citet{h20}. To focus on effects of the initial vertical distribution of particles, we assume that the particles are initially on circular orbits. This assumption is valid for the case of $r_{\mr{d}}\leq r_{\mr{d,crit}}$, because eccentricities of such small particles are quickly damped by gas drag \citep[see  Figure 11 in][]{h20}.
Then, the parameters that determine the initial conditions of particles are the impact parameter $b=a_{\mr{s}}-a$ and the initial altitude $z_{\mr{0}}$, where $a_{\mr{s}}$ is the initial semi-major axis of the particle. We start orbital integration at the azimuthal boundary of the hydrodynamic simulation box; $x>0$ and $y_0=L_y/2 - L_y/n_y$ \citep{h20}. We derive the initial $x$-coordinate from $y_0$ and the given impact parameter $b$ as $x_0=\sqrt{b^2-8/y_0}$ \citep{in89}.
While large particles that are not well-coupled to the gas can be trapped at the pressure bump of the gas gap \citep{z12,w18}, we consider those particles that enter the gap by some mechanisms and assume that the radial distribution of these particles within the gap is uniform. The actual surface density of particles within the gap should depend on their Stokes number;  
we will discuss dust accretion into the CPD considering the dust gap based on results of global simulation in \Secref{sec:appendix}.

The conditions for the capture of particles and the termination of the orbital integration are the same as in \citet{h20} \citep[see also][]{t14}.
Each particle is considered to be captured by the CPD when one of the following two conditions is met:
(i) The orbit of an energetically captured particle becomes planet-centered and nearly circular in the CPD: $\tilde{E}<0,\; e_{\mr{p}}<0.3,\; \tilde{a}_{\mr{p}}<0.5,\; |N_{\mr{w}}|\geq 3$.
Here $\tilde{E}$ is the energy integral given by
\begin{equation}
\tilde{E}=\frac{1}{2}(\dot{\tilde{x}}^2+\dot{\tilde{y}}^2+\dot{\tilde{z}}^2) - \frac{1}{2}(3\tilde{x}^2 - \tilde{z}^2) - \frac{3}{\tilde{r}} + \frac{9}{2}.
\end{equation}
In the above, $e_{\mr{p}}$ and $a_{\mr{p}}$ are the eccentricity and semi-major axis of the planet-centered orbit of the particle, respectively, and are given by \citep{md99}:
\begin{eqnarray}
\tilde{a}_{\mr{p}} &=& \left( \frac{2}{\tilde{r}} -\frac{\tilde{v}_{\mr{iner}}^2}{3} \right)^{-1}, \\
e_{\mr{p}} &=& \sqrt{1-\frac{j_{\mr{iner}}^2}{3a_{\mr{p}}}},
\end{eqnarray}
where $v_{\mr{iner}}$ and $j_{\mr{iner}}$ are the velocity and the specific angular momentum around the planet in the inertial frame, respectively.
$N_{\mr{w}}$ is the winding number around the planet \citep{io07}.
(ii) Even if the particle is not in a circular orbit around the planet, it is considered to be captured if it is captured energetically within the planet's Hill sphere and has orbited around the planet a sufficient number of times: $\tilde{E}<0$, $\tilde{r}<1$, $|N_{\mr{w}}|\geq 15$.
\footnote{While we consider the capture of particles in both the prograde and retrograde directions, we find that all the particles captured in the prograde direction because the size of particles is sufficiently small and particles are dragged by the gas flow in the CPD.}

We terminate the orbital integration for each particle when any one of the above capture conditions (i) or (ii), or either of the following conditions is met:
(iii) the particle collides with the planet: $r<r_{\mr{p}}$, where $r_{\mr{p}}$ is the physical radius of the planet. We set $r_{\mr{p}}=10^{-3}r_{\mr{H}}$ assuming the Jovian orbit \citep{in89,o12}.
(iv) The particle is far enough away from the planet: $r>y_0 + b + (2e_0 + i_0)a_0$, where $e_0$, $i_0$, and $a_0$ are the eccentricity, inclination, and semi-major axis of the particle's initial heliocentric orbit, respectively.

%% file: 03result.tex

\subsection{Examples of Particle Orbits}\label{sec:orbit_ex}
The orbital behavior of particles from the PPD to the CPD depends on their Stokes number $St$ \citep{h20}. 
Here we define $St$ by the following equations using the planet-centered-Keplerian angular velocity $\Omega_{\mr{p}}$ near the planet ($\tilde{r}<0.5$) and star-centered-Keplerian angular velocity $\Omega_{\mr{K}}$ elsewhere:
\begin{equation}\label{eq:stokes}
St \equiv
\begin{dcases}
t_{\mr{stop}}\Omega_{\mr{p}} \qquad (\mr{for} \quad \tilde{r}< 0.5), \\
t_{\mr{stop}}\Omega_{\mr{K}} \qquad (\mr{for} \quad \tilde{r}\geq 0.5),
\end{dcases}
\end{equation}
where the stopping time $t_{\mr{stop}}$ is described as
\begin{equation}
t_{\mr{stop}}=m_{\mr{d}}\Delta u/F_{\mr{drag}}=\frac{8\rho_{\mr{d}}r_{\mr{d}}}{3C_{\mr{D}}\rho_{\mr{g}}\Delta u}.
\end{equation}

\begin{figure}[H]

\begin{tabular}{c}
\vspace{1cm}
	\begin{minipage}{0.5\hsize}	
		\begin{center}
			\includegraphics*[bb=0 0 336 310,scale=0.9]{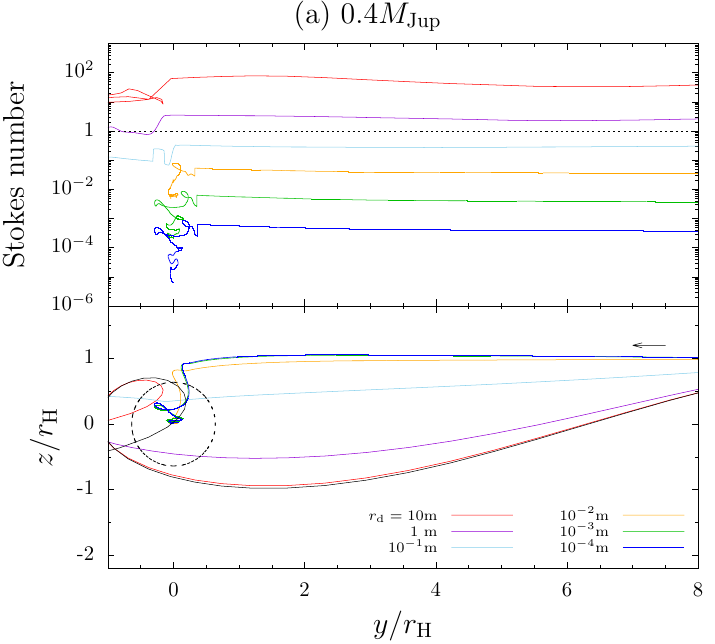}
		\end{center}
	\end{minipage}\\

	\begin{minipage}{0.5\hsize}	
		\begin{center}
			\includegraphics*[bb=0 0 336 310,scale=0.9]{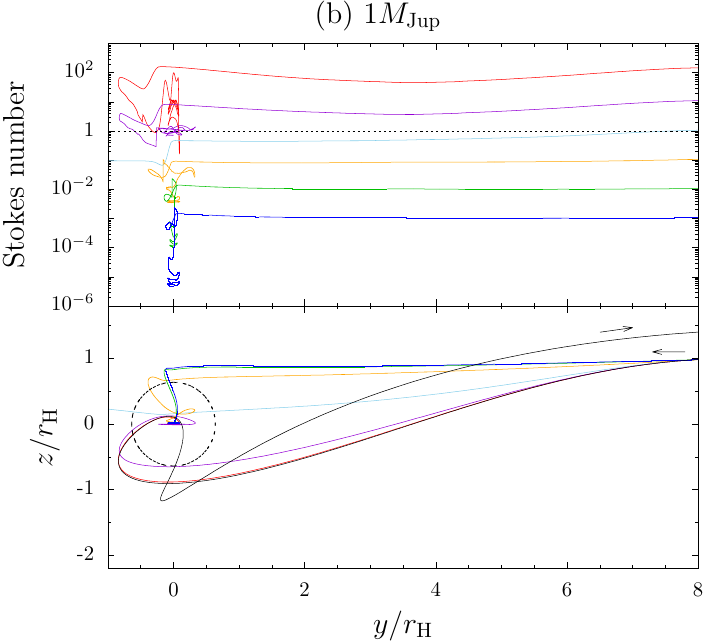}
		\end{center}
	\end{minipage}

\end{tabular}

	\caption{Examples of particle orbits and change of the Stokes number along the orbits in the case of $(\tilde{b},\tilde{z}_{\mr{s}})=(2.3,\; 1.0)$ and $(2.4,\; 1.0)$ for $M_{\mr{p}}=0.4M_{\mr{Jup}}$ (top panel) and $M_{\mr{p}}=1M_{\mr{Jup}}$ (bottom panel), respectively. 
The colors show the size of particles; red, violet, light blue, orange, green, and blue correspond to $r_{\mr{d}}=10$, 1, 0.1, $10^{-2}$, $10^{-3}$, and $10^{-4}$~m, respectively. The black lines show the case without gas drag. The dashed lines in the bottom panels show the Hill sphere of the planet.}

	\label{fig:orbit_st}
\end{figure}

\Figref{fig:orbit_st} shows examples of particle orbits and the change of the Stokes number along them in the case of $z_{\mr{s}}=r_{\mr{H}}$.
In the case of $r_{\mr{d}}=10$~m (red), $St\gg 1$ and the effect of gas drag is sufficiently small, and the orbit is similar to that without gas drag (black).
It oscillates in the vertical direction with the amplitude depending on the initial altitude and can be captured by the CPD when it happens to pass close to the planet.

In the cases of $0.1\leq r_{\mr{d}} \leq 1$~m (light blue and violet), $St\sim 1$ is maintained while approaching the Hill sphere. In the case of $r_{\mr{d}}=0.1$~m $\simeq r_{\mr{d,crit}}$, $St=0.5$ and the particle gradually settles toward the mid-plane.
When $St<0.5$, particles cannot settle completely and approach the Hill sphere at an altitude slightly above the mid-plane and are captured from above by the CPD.
On the other hand, when $St\gtrsim 0.5-1$, the vertical oscillation is not completely damped, and the particles approach the Hill sphere with a slight oscillation \citep[see also][]{h20}.
The degree of oscillation damping depends on the density of the gas along the orbit, i.e., the depth of the gap formed around the planet's orbit.
It should be noted that the gap depth obtained by our local simulation tends to be shallower than the one obtained by global simulation for the same planetary mass.
Since the capture process for small particles with $St\lesssim 1$ are sensitive to the depth of the gap, it is important to take such effects into account when examining planetary-mass dependence \citep[][see also \Secref{sec:caution}]{m22}.

The particles with $r_{\mr{d}} \leq 1$~cm (orange, green, and blue) have $St \ll 1$ and approach the Hill sphere while keeping a certain altitude, being more or less coupled to the gas flow. Then they move down nearly vertically with the accreting gas flow over the Hill sphere and become captured by the CPD.

Differences in the gas field around a planet due to differences in planetary mass affect particle orbits in the following ways: First, the impact parameter $b$ of the accretion orbit is varied because of the structure of the accretion bands is modified (see \Secref{sec:bzcap}). 
Second, when the planetary mass is increased, the gas density around the planetary orbit tends to be reduced more due to gap opening, resulting in a larger Stokes number for the same particle size. This makes the particles with $St \simeq 1$ ($r_{\rm d}=10^{-1}$~m) more easily settled toward the midplane while approaching the planet.

\subsection{Initial Orbits of Captured Particles}\label{sec:bzcap}
\begin{figure}[H]
	\begin{center}
		\includegraphics*[bb=0 0 390 220,scale=1]{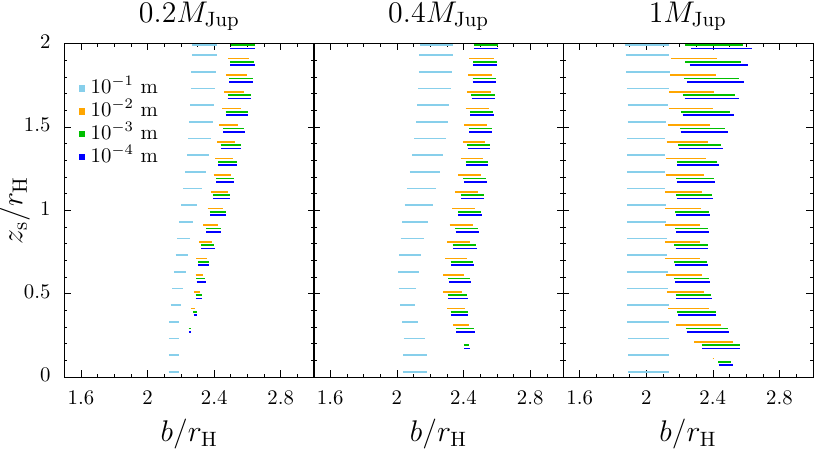}
	\end{center}
	\caption{Distribution of the initial orbital parameters $(b,z_{\mr{s}})$ of captured particles (dust accretion band). The colors show the size of the particle; light blue, orange, green, blue corresponds to $r_{\mr{d}}=0.1$, $10^{-2}$, $10^{-3}$, and $10^{-4}$~m, respectively. The vertical axis is plotted at every $z_{\mr{s}}=0.1r_{\mr{H}}$, but slightly shifted up and down to avoid overlapping of the lines.}
	\label{fig:bzcap}
\end{figure}

In \Figref{fig:bzcap}, we show the distribution of the initial impact parameter $b$ and initial altitude $z_{\mr{s}}$ of the particles captured by the CPD.
The different colors mean the different sizes of the particles. The results for $r_{\rm d}=0.1$~cm (light blue) are shown as examples for the case of $St\sim 1$ for comparison with other cases of $St\ll 1$.
In the case of the particles with $St\ll 1$ ($r_{\rm d}\leq 10^{-2}$~m), the initial orbits of captured particles distribute above the midplane continuously in the $z$-direction in a zonal pattern except for $z_{\mr{s}}\simeq 0$.
Because particles of this size are greatly affected by the gas flow, the distribution of the dust accretion band is very similar to that of the gas accretion band \citep{t12,m22}.
We confirm that sufficiently small particles initially near the midplane of the PPD cannot accrete into the CPD due to the radially outward gas flow near the midplane in the Hill sphere \citep{t14,h20,m22}.

In the case of $St \ll 1$ ($r_{\rm d}\leq 10^{-2}$~m), the radial width of the dust accretion band measured in units of $r_{\rm H}$ becomes wider with increasing planetary mass.
For example, in the case of $r_{\mr{d}}= 0.1$~mm (blue), while the width of the dust accretion band is $\Delta b_{\mr{cap}} \lesssim 0.15r_{\mr{H}}$ for $M_{\mr{p}}=0.4M_{\mr{Jup}}$, $\Delta b_{\mr{cap}} \gtrsim 0.3r_{\mr{H}}$ at maximum for $M_{\mr{p}}=1M_{\mr{Jup}}$.
This expansion of $\Delta b_{\mr{cap}}$ is pronounced in the case of smaller particles.
This is because the smaller particles are more affected by the gas flow and the width of the gas accretion band is wider for larger planetary masses \citep{m22}.
In the case of $M_{\rm p}=1M_{\rm Jup}$, $\Delta b_{\mr{cap}}$ is larger at $z_{\rm s}\gtrsim r_{\rm H}$, which is attributed to the fact that the gas accretion band becomes slightly wider at $z\gtrsim r_{\rm H}$ \citep[see Fig.7 in][]{m22}.
Also, $\Delta b_{\mr{cap}}$ is almost independent of $z_{\rm s}$ when $r_{\rm d}=0.1$~m because the particles settle toward the midplane before they approach the planet. Thus, the range of $b$ that leads to accretion is similar to that of $z_{\rm s} \sim 0$, regardless of initial altitudes.

\subsection{Distribution of the Locations of Particle Capture in the CPD}\label{sec:cap-hist}
\begin{figure}[H]
	\begin{center}
		\includegraphics*[bb=0 0 432 432,scale=0.8]{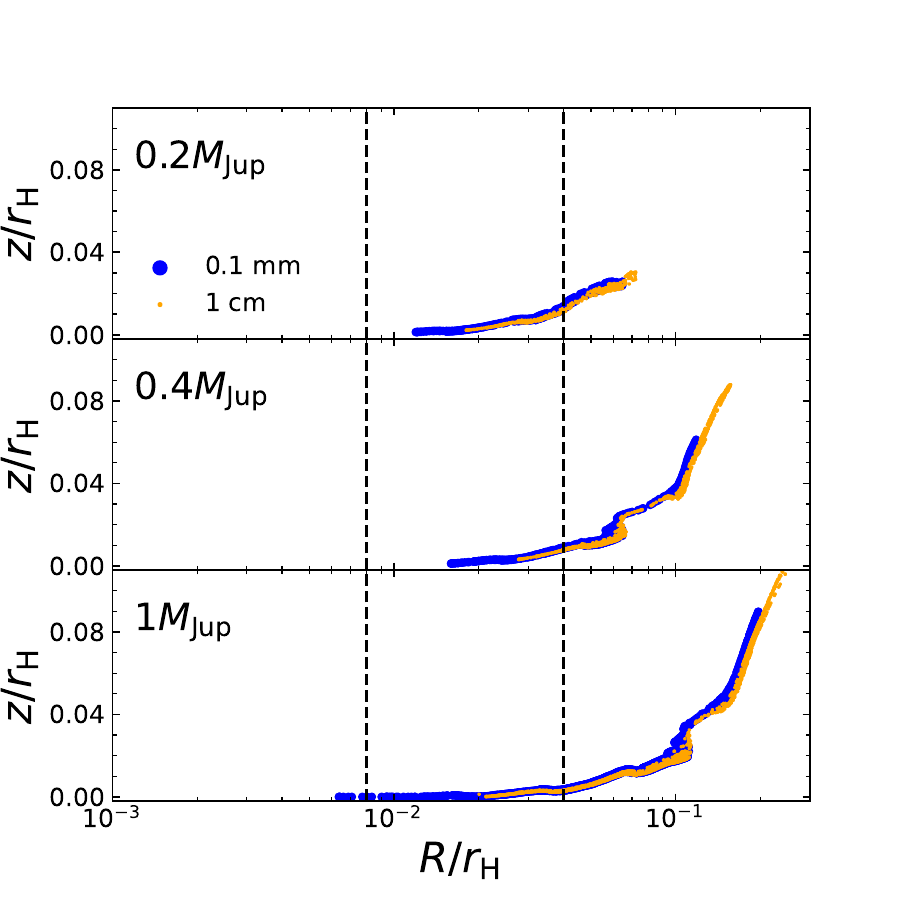}
	\end{center}
	\caption{Radial and vertical positions $(R,\; z)$ where the particle capture conditions are met for the case of $0<z_{\mr{s}}\leq r_{\mr{H}}$. Orange and blue points represent results for $r_{\mr{d}}=10^{-2}$~m and $10^{-4}$~m, respectively. The two black vertical dashed lines show the current radial positions of Io and Callisto. The results for $r_{\mr{d}}=10^{-3}$~m are not shown here, but are similar to the case of $r_{\mr{d}}=10^{-4}$~m.}
	\label{fig:cap-hist}
\end{figure}

\Figref{fig:cap-hist} shows the distribution of the coordinates $(R,\; z)$ of particles when their capture conditions are met in the range $0<z_{\mr{s}}\leq r_{\mr{H}}$ for the cases of two different sizes. We show the results for only two cases of $r_{\mr{d}}=10^{-2}$~m and $10^{-4}$~m. The results for $r_{\mr{d}}=10^{-3}$~m are similar to the case of $r_{\mr{d}}=10^{-4}$~m and are not shown here.
In agreement with \citet{h20}, particles accreting with the gas are widely supplied onto the surface of the CPD.
We found that the particles with $r_{\mr{d}}=10^{-4}$~m tend to be captured at more inner region of the CPD than those with $10^{-2}$~m.
This suggests that small particles accreting from high altitudes can be delivered into more inner regions of the CPD with the gas \citep{h20}.
On the other hand, if particles in the PPD undergo significant settling in approaching the planet, they are captured in the midplane near the outer edge of the CPD.
Although the size of the CPD is commonly $r_{\rm CPD}\simeq 0.2 r_{\rm H}$ when $r_{\rm H}/h_{\rm g}\gtrsim 1$ \citep[e.g., Figure 4 in][]{m22}, we found that the Hill-scaled radial positions of particle captured on the CPD are more widely distributed in the case of larger planets.
This may reflect the fact that the size of the CPD measured in units of $r_{\rm H}$ and/or the gas density near the outer edge of the CPD is somewhat larger for a larger planetary mass \citep[Figures 3 and 4 in][]{m22}.

\subsection{Capture Rate of Dust Particles onto the CPD}\label{sec:rate_z}
Next, we calculate the capture rates of particles onto the CPD.
The capture rates in this work are obtained by counting the number of captured particles per unit time and unit surface number density of particles in the PPD, and $P_{\mr{cap}}$ is normalized by $r_{\rm H}^{2}\Omega_{\rm K}$.
For $r_{\mr{d}}<r_{\mr{d,crit}}$, the capture rate of particles which have initial altitude $z_{\mr{s}}$ is described by \citep{f13,t14,h20}:
\begin{equation}\label{eq:rate-z}
P_{\mr{cap}}(r_{\mr{d}},\; z_{\mr{s}})= \int^{\infty}_{-\infty} \varphi_{\mr{cap}} (r_{\mr{d}},\; z_{\mr{s}},\; \tilde{b})\frac{3}{2}|\tilde{b}|d\tilde{b},
\end{equation}
where $\varphi_{\mr{cap}}$ is defined such that $\varphi_{\mr{cap}}=1$ for capture orbits and $\varphi_{\mr{cap}}=0$ otherwise. We adopted $d\tilde{b}=5\times 10^{-4}$ and performed calculations by varying $z_{\rm s}$ in increments of 0.1.

\begin{figure}[H]
		\begin{center}
			\includegraphics*[bb=0 0 339 242,scale=0.8]{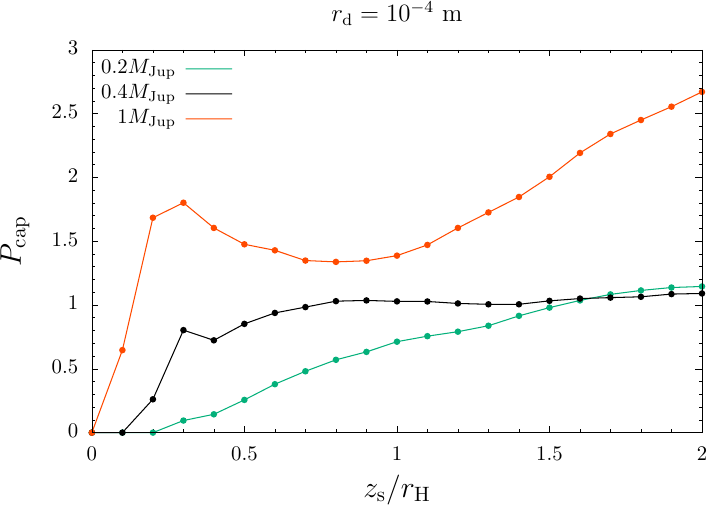}
		\end{center}
		\caption{Capture rate $P_{\mr{cap}}$ as a function of the initial altitude $z_{\rm s}$ for $r_{\mr{d}}=10^{-4}$~m. The colors show the difference of the planetary mass; green, black, red correspond to $M_{\mr{p}}=0.2M_{\mr{Jup}}$, $0.4M_{\mr{Jup}}$, and $1M_{\mr{Jup}}$, respectively.}
	\label{fig:rate_z_r-4}
\end{figure}

\Figref{fig:rate_z_r-4} shows the capture rates for $r_{\mr{d}}=10^{-4}$~m. 
Although $P_{\mr{cap}}$ is non-dimensionalized by the Hill-scaling, the overall values of $P_{\mr{cap}}$ are larger for larger planetary masses, reflecting the increase in the width of the gas accretion band scaled by $r_{\rm H}$ (\Figref{fig:bzcap}).
The increase of $P_{\mr{cap}}$ at $z_{\rm s}\gtrsim r_{\rm H}$ for $M_{\rm p}=1M_{\rm Jup}$ is attributed to the fact that the gas accretion band becomes slightly wider in this altitude range, as mentioned in \Secref{sec:bzcap}.
We also found that $P_{\mr{cap}}=0$ at $z_{\mr{s}}=0$ for all the three cases, owing to the outward gas flow near the midplane in the Hill sphere (\Secref{sec:bzcap}).
The increase of the capture rate of dust with small $z_{\rm s}$ $(\lesssim 0.5 r_{\rm H})$ for $M_{\rm p}=1M_{\rm Jup}$ can be explained by the extension of the gas accretion band to lower altitudes \citep{m22}.

\begin{figure}[H]

\begin{tabular}{cc}
\hspace{-1cm}
	\begin{minipage}{0.5\hsize}	
		\begin{center}
			{\includegraphics*[bb=0 0 339 242,scale=0.7]{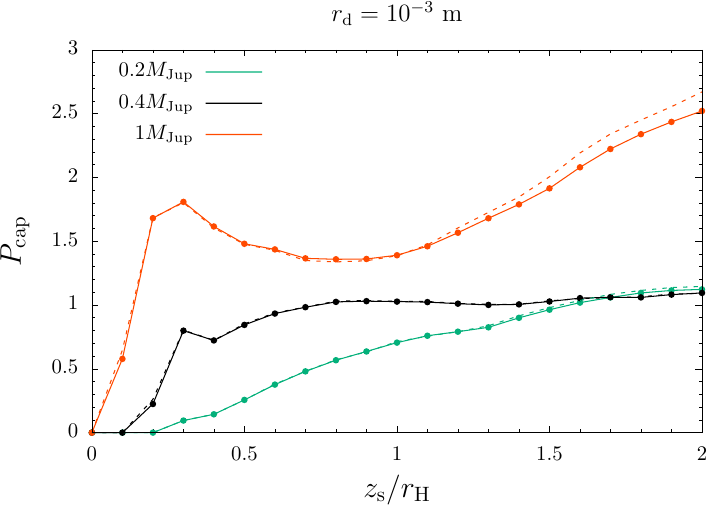}}
		\end{center}
	\end{minipage}&
	
	\begin{minipage}{0.5\hsize}	
		\begin{center}
			\includegraphics*[bb=0 0 339 242,scale=0.7]{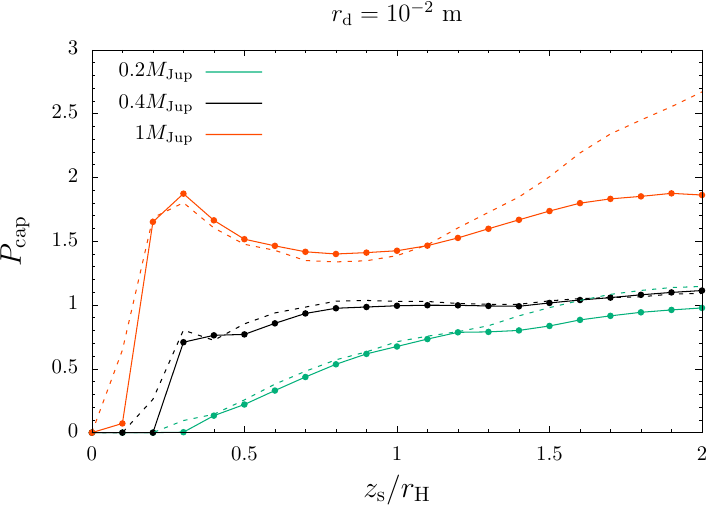}
		\end{center}
	\end{minipage}

\end{tabular}
	\caption{Same as \Figref{fig:rate_z_r-4} but for $r_{\mr{d}}=10^{-3}$~m (left panel) and $10^{-2}$~m (right panel), respectively. The dashed line shows the results for $r_{\mr{d}}=10^{-4}$~m for comparison.}
	\label{fig:rate_z_r-1-3}
\end{figure}

Results for larger particle sizes are shown in \Figref{fig:rate_z_r-1-3}, with those for $r_{\mr{d}}=10^{-4}$~m with the dashed lines for comparison. 
We find that the results are quite similar for all the three particle sizes, because in almost all these cases dust particles have sufficiently small Stokes numbers and are coupled well with the gas.
As an exception, a significant deviation from the other cases can be seen in the case of $r_{\mr{d}}=10^{-2}$~m at $z_{\mr{s}} \gtrsim 1.5h_{\rm g}$ when $M_{\rm p}=1M_{\rm Jup}$. This is because the gas density is very low at such high altitudes ($z_{\mr{s}} \gtrsim 2h_{\rm g}$), so the Stokes number of the particles is large.

%% file: 04discussion.tex
\subsection{Capture Rates of Vertically Stirred Dust Particles}\label{sec:Pcap-total}
Integrating the capture rates per unit surface number density of particles in the PPD (i.e., $P_{\rm cap}(r_{\rm d},\; z_{\rm s})$ given by \Equref{eq:rate-z}) over $z_{\rm s}$ with a given vertical distribution of particles, we can calculate the total capture rates of particles into the CPD.
The degree of vertical stirring of dust particles in the PPD can be expressed by the dust scale height $h_{\rm d}$. \citet{h20} considered dust stirring due to turbulence and used the dust scale height $h_{\rm d}$ expressed in terms of the turbulent viscosity parameter $\alpha$ \citep{ss73} and the Stokes number of particles $St$ \citep{yl07}. Then they obtained total capture rates and mass accretion rates of dust particles for several combinations of $\alpha$ and $St$.
However, turbulence is not the only mechanism that stirs up dust particles in the PPD.
For example, recent studies show that the meridional circulation caused by an embedded planet can stir up dust particles \citep{bi21,sz22}.

\begin{figure}[H]
\begin{tabular}{c}
	\begin{minipage}{1\hsize}	
		\begin{center}
			\includegraphics*[bb=0 0 365 260,scale=0.8]{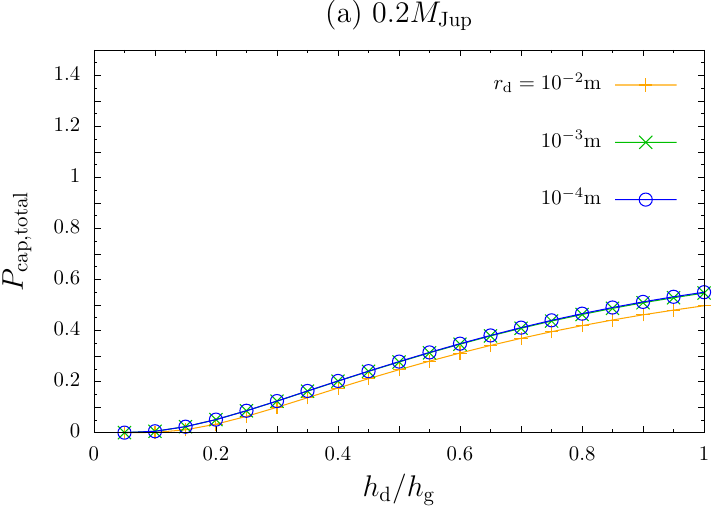}
		\end{center}
	\end{minipage}\\
	
\vspace{0.5cm}
	\begin{minipage}{1\hsize}	
		\begin{center}
			\includegraphics*[bb=0 0 365 260,scale=0.8]{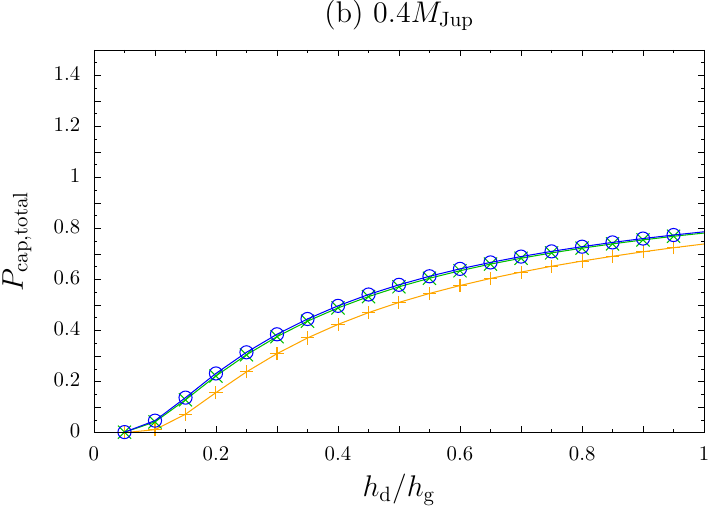}
		\end{center}
	\end{minipage}\\
	
\vspace{0.5cm}
	\begin{minipage}{1\hsize}	
		\begin{center}
			\includegraphics*[bb=0 0 365 260,scale=0.8]{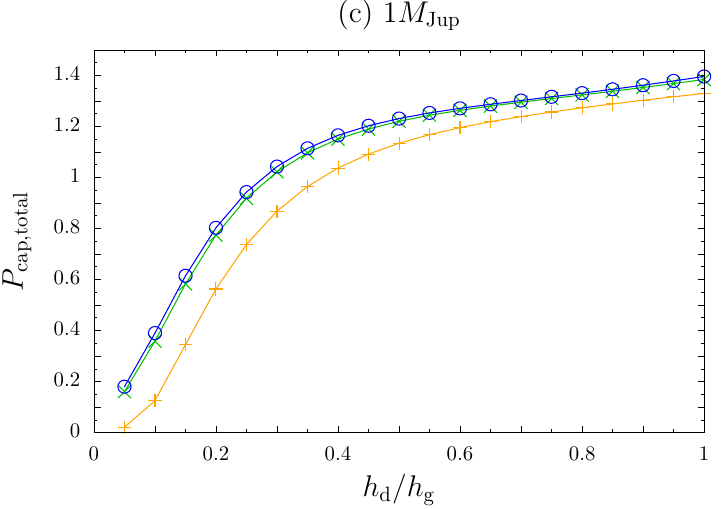}
		\end{center}
	\end{minipage}

\end{tabular}
	\caption{Total capture rate $P_{\rm cap,total}$ as a function of $h_{\rm d}/h_{\rm g}$. Top, middle, and bottom panels show the results for $M_{\rm p}=0.2$, 0.4, and $1M_{\rm Jup}$, respectively. Colors show the dust size, and orange, green, and blue correspond to $r_{\rm d}=10^{-2}$, $10^{-3}$, and $10^{-4}$~m, respectively.}
	\label{fig:Ptotal-hd}
\end{figure}

Therefore, rather than focusing on any specific mechanism for dust particle stirring, in the present work we will obtain the total capture rate $P_{\mr{cap,total}}$, treating the dust scale height $h_{\rm d}$ as a parameter.
We assume that the vertical distribution of particles is given by the Gaussian distribution with a scale height $h_{\rm d}$;
\begin{equation}\label{eq:n}
n(h_{\rm d}, \: z_{\mr{s}}) = \frac{\Sigma_{\mr{d}}}{\sqrt{2\pi}h_{\mr{d}}m_{\mr{d}}} \exp \left( -\frac{z_{\mr{s}}^2}{2h^2_{\mr{d}}} \right).
\end{equation}
Then the total capture rate of particles $P_{\mr{cap,total}}$ with this vertical distribution is given by
\begin{equation}
P_{\mr{cap,total}}(h_{\rm d})=\frac{\int^{\infty}_{-\infty}P_{\mr{cap}}(z_{\mr{s}})n(h_{\rm d},\: z_{\mr{s}})dz_{\mr{s}}}{\int^{\infty}_{-\infty}n(h_{\rm d},\: z_{\mr{s}})dz_{\mr{s}}}.
\end{equation}
\Figref{fig:Ptotal-hd} shows the plots of $P_{\mr{cap,total}}(h_{\rm d})$. The behavior of $P_{\mr{cap,total}}(h_{\rm d})$ can be roughly explained by the $z_{\rm s}$-dependence of the width of the dust accretion band (\Figref{fig:bzcap}), i.e., the total capture rate is larger for a larger planetary mass and a larger dust scale height.

\citet{bi21} investigated effects of planet-driven meridional circulation on vertical stirring of dust using three-dimensional hydrodynamic simulation considering the back-reaction of dust on the gas. They showed that even in the case of a planet with a Saturnian mass ($M_{\rm p}\simeq 0.3M_{\rm Jup}$), dust particles are easily stirred up to $h_{\rm d}\simeq 0.6 h_{\rm g}$.
In the case of $h_{\rm d}= 0.6 h_{\rm g}$ in \Figref{fig:Ptotal-hd}, we find that $P_{\mr{cap,total}} \simeq 0.4$ for $M_{\rm p}=0.2M_{\rm Jup}$, $\simeq 0.6$ for $M_{\rm p}=0.4M_{\rm Jup}$, and $\simeq 1.2$ for $M_{\rm p}=1M_{\rm Jup}$, respectively.
That is, although $P_{\mr{cap,total}}$ is normalized by $r_{\rm H}^2 \Omega_{\rm K}$ as mentioned above, its value for $M_{\rm p}=1M_{\rm Jup}$ is about three times larger than for $M_{\rm p}=0.2M_{\rm Jup}$ as compared to the simple Hill-scaling.
As for the dependence on particle size in the size range considered here (0.1~mm ― 1~cm), the total capture rate is roughly determined by the dust scale height and the planetary mass, suggesting that the particle size dependence is not significant.
However, since the dust scale height likely depends on the particle size (or the Stokes number), the total capture rate likely depends on the dust size via the dust scale height.

\subsection{Degree of Dust Retention in Accreting Gas}\label{sec:dd}
Recent theoretical models of satellite formation in the CPD suggest that one of the key parameters for the formation of satellitesimals and satellites in the CPD is the ratio of dust-to-gas inflow rates \citep{sh17,sh19}. Using $P_{\mr{cap,total}}$ we obtained above, we can derive the mass accretion rate of dust into the CPD from \citep{h20}
\begin{equation}\label{eq:Macc}
\dot{M}_{\mr{d}} = \Sigma_{\mr{d}}r_{\mr{H}}^2 \Omega_{\mr{K}}P_{\mr{cap,total}},
\end{equation}
where $\Sigma_{\rm d}$ is the dust surface density at the position of the dust accretion band.

Now, following \citet{m22}, we consider the gas flowing through the area of the accretion band at $(x,y)=(x_{\rm g,ab},\; L_{y}/2)$, where $x_{\rm g,ab}$ is a typical value of the $x$-coordinate of the accretion band. The velocity of the gas passing through the area of the accretion band is $(3/2)x_{\rm g, ab}\Omega_{\rm K}$ in the coordinate system rorating with the planet. Then, the mass accretion rate of the gas into the CPD is expressed as:
\begin{equation}\label{eq:mdot_g}
\dot{M}_{\rm g} = D_{\rm g} \Sigma_{\rm g},
\end{equation}
\begin{equation}\label{eq:d_g}
D_{\rm g}= 2\times \frac{3}{2}x_{\rm g,ab} \Omega_{\rm K} \bar{w}_{\rm g},
\end{equation}
where $D_{\rm g}$ is the \textit{accretion area} of the gas \citep{tt16}. 
We obtain $\bar{w}_{\rm g}$ by averaging the width of the accretion band $w_{\rm g}(z)$ over $z$ with the gas density distribution \citep{m22}:
\begin{equation}\label{eq:bar_w_g}
\bar{w}_{\rm g}=\frac{\int^{z_{\mr{max}}}_0 w_{\rm g}(z)\exp \left( -\frac{z^2}{2h_{\rm{g}}^2} \right)dz }{\int^{z_{\mr{max}}}_0 \exp \left( -\frac{z^2}{2h_{\rm{g}}^2} \right)dz}.
\end{equation}

Similarly, the dust accretion rate into the CPD can be obtained using the accretion area of the dust particles $D_{\rm d}$ and the averaged width of the dust accretion band $\bar{w}_{\rm d}$ as
\begin{equation}\label{eq:mdot_d}
\dot{M}_{\rm d} = D_{\rm d} \Sigma_{\rm d},
\end{equation}
\begin{equation}\label{eq:d_d}
D_{\rm d}= 2\times \frac{3}{2}x_{\rm d,ab} \Omega_{\rm K} \bar{w}_{\rm d},
\end{equation}
\begin{equation}\label{eq:bar_w_d}
\bar{w}_{\rm d}=\frac{\int^{z_{\mr{max}}}_0 w_{\rm d}(z)\exp \left( -\frac{z^2}{2h_{\rm{d}}^2} \right)dz }{\int^{z_{\mr{max}}}_0 \exp \left( -\frac{z^2}{2h_{\rm{d}}^2} \right)dz}.
\end{equation}
Then the ratio of dust-to-gas inflow rates $\chi$ is given by
\begin{equation}\label{eq:chi}
\chi = \frac{\dot{M}_{\rm d}}{\dot{M}_{\rm g}}=\frac{D_{\rm d} \Sigma_{\rm d}}{D_{\rm g} \Sigma_{\rm g}}.
\end{equation}
In the limit of a very small dust size, dust particles are completely coupled to the gas and we have $h_{\rm d}=h_{\rm g}$ and $\bar{w}_{\rm d}=\bar{w}_{\rm g}$, thus $D_{\rm d}=D_{\rm g}$.
In this case, we obtain $\chi=\Sigma_{\rm d}/\Sigma_{\rm g}$, i.e., the ratio of dust-to-gas inflow rates is determined by the ratio of dust-to-gas surface density at the accretion band (feeding zone).
On the other hand, if the dust-gas coupling is not perfect, $h_{\rm d}<h_{\rm g}$ due to dust settling, and dust is depleted in the gas accreting onto the CPD, resulting in $D_{\rm d}/D_{\rm g}<1$ \citep{t12}.
Here, we call $D_{\rm d}/D_{\rm g}$ as \textit{degree of dust retention} of accreting gas into the CPD.

The surface densities of dust and gas at the accretion band depend on the depth and width of the gap in the PPD, which depends on various parameters, such as the viscosity of the PPD.
In order to avoid uncertainty related to such uncertain parameters, we will focus on the degree of dust retention $D_{\rm d}/D_{\rm g}$ to examine the planetary mass dependence of the ratio of dust-to-gas inflow rates in the accretion band.
For the range of planetary masses considered in this study corresponding to $0.8 \leq r_{\rm H}/h_{\rm g} \leq 1.36$, the width of the gas accretion band is approximately given by \citep{m22}
\begin{equation}\label{eq:w_g_ana}
\frac{\bar{w}_{\rm g}}{h_{\rm g}} = 0.12 \left( \frac{r_{\rm H}}{h_{\rm g}} \right)^3.
\end{equation}
Substituting \Equref{eq:w_g_ana} into \Equref{eq:d_g} and setting $x_{\rm g,ab}=2.2r_{\rm H}$, we obtain
\begin{equation}\label{D_g_ana}
D_{\rm g} = 0.79 r_{\rm H}^4 h_{\rm g}^{-2} \Omega_{\rm K}.
\end{equation}
As for the accretion of dust, from \Equrefs{eq:Macc} and (\ref{eq:mdot_d}), we obtain
\begin{equation}\label{eq:D_d_ana}
D_{\rm d} = P_{\rm cap,total} r_{\rm H}^2 \Omega_{\rm K}.
\end{equation}
From \Equrefs{D_g_ana} and (\ref{eq:D_d_ana}), we can obtain the degree of dust retention of accreting gas into the CPD as
\begin{equation}\label{eq:Dd-Dg}
\frac{D_{\rm d}}{D_{\rm g}} = 1.26 P_{\rm cap,total} \left( \frac{r_{\rm H}}{h_{\rm g}} \right)^{-2},
\end{equation}
where $P_{\rm cap,total}$ can be obtained from our numerical results. 

\begin{figure}[H]	
	\begin{center}
		\includegraphics*[bb=0 0 203 290,scale=1.0]{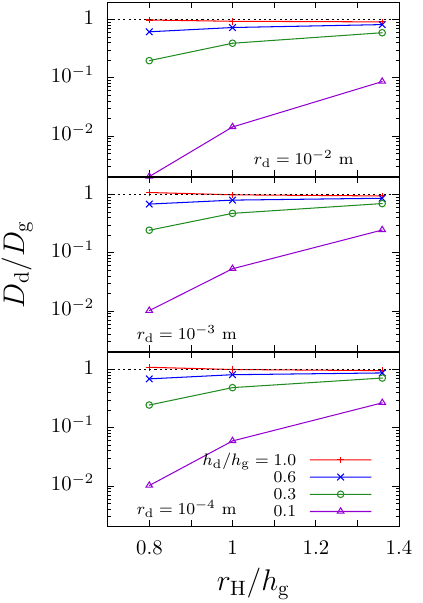}
	\end{center}
	
	\caption{Degree of dust retention of accreting gas into the CPD as a function of $r_{\rm H}/h_{\rm g}$. The top, middle, bottom panels show the case of $r_{\rm d}=10^{-2}$, $10^{-3}$, and $10^{-4}$~m, respectively. Color means the dust scale height in the PPD; violet, green, blue, and red correspond to $h_{\rm d}/h_{\rm g}=0.1$, 0.3, 0.6, and 1.0, respectively.}
	\label{fig:dd-mp}
\end{figure}

\Figref{fig:dd-mp} shows the plots of $D_{\rm d}/D_{\rm g}$ obtained from \Equref{eq:Dd-Dg}.
We confirm that $D_{\rm d}/D_{\rm g}\simeq 1$ when dust particles are perfectly coupled to the gas and $h_{\rm d}/h_{\rm g}\simeq 1$, while $D_{\rm d}/D_{\rm g}$ decreases with decreasing dust scale height due to insufficient dust stirring in the PPD.
The decrease in $D_{\rm d}/D_{\rm g}$ at small dust scale heights is more pronounced for smaller planetary masses, because the accretion band of dust and gas becomes narrower at low altitudes for small planetary masses (\Figref{fig:bzcap}).
On the other hand, for $r_{\rm H}/h_{\rm g}=1.36$ ($M_{\rm p}=1M_{\rm Jup}$), $D_{\rm d}/D_{\rm g}\gtrsim 0.1$ even when $h_{\rm d}/h_{\rm g}=0.1$.

When the surface densities of dust and gas at the accretion band are given, we can determine the ratio of dust-to-gas inflow rates into the CPD combining them with the degree of dust retention $D_{\rm d}/D_{\rm g}$ obtained above.
The ratio of surface densities of dust and gas at the accretion band depends on the depth and width of gap opened by the planet and the radial position of the accretion band relative to the gap.
As an example, in \Secref{sec:appendix}, we derive dust mass accretion rates onto the CPD and the ratio of dust-to-gas inflow rates in the case where the surface densities of dust and gas at the accretion band are given using simple models based on results of global hydrodynamic simulation.

The surface density of dust at the accretion band depends on dust size.
Particles with $St \sim 1$ are easily trapped at the pressure bump of the gap edge and the surface density becomes high \citep{z12,w18}, and even planetesimals may form there \citep{e20,sa20,sa23}. Then collision and fragmentation of the planetesimals at the pressure bump can provide a significant amount of dust, which can diffuse into the gap \citep{k12,st23}.
While recent studies using hydrodynamic simulations including gas and dust components obtain empirical formulae for the depth and width of the dust gap \citep{z18}, detailed studies of dynamical behavior of dust and gas around the gap is still in progress.
Further investigation of the surface density distribution of dust and gas around the planetary orbit will be important to obtain the ratio of dust-to-gas inflow rates into the CPD and derive constraints on satellite formation.

\subsection{Applications}\label{sec:appendix}
\subsubsection{Dust Mass Accretion Rates}\label{ap:mdot}
In order to obtain dust mass accretion rates onto the CPD using \Equref{eq:Macc} with our numerical results for $P_{\rm cap,total}$, we need to assume the dust surface density in the accretion band. Here, we derive dust mass accretion rates using the surface density at the bottom of the gap.

\citet{z18} performed two-dimensional two-fluid hydrodynamic simulations for dust and gas, and obtained empirical formulae for the width and depth of the dust gap.
The formulae obtained for the depth of the dust gap is
\begin{eqnarray}\label{eq:zhang18}
\label{eq:zhang18_Sd} \frac{\Sigma_{\rm d,gap}}{\Sigma_{\rm d,peak}} &=& \frac{1}{1+CK^D},\\
\label{eq:zhang18_K} \frac{K}{24}&=&\frac{M_{\rm p}/M_{\rm c}}{0.001}\left( \frac{h_{\rm g}/a}{0.07} \right)^{-2.81} \left(\frac{\alpha_{\rm PPD}}{10^{-3}} \right)^{-0.38},
\end{eqnarray}
where $\Sigma_{\rm d,gap}$ and $\Sigma_{\rm d,peak}$ are the bottom and peak values of the dust surface density distribution, respectively. The values of the fitting parameters $C$ and $D$ for various values of the Stokes number of the dust particles are obtained from the hydrodynamic simulations \citep[see Table 2 in][]{z18}.
We assume $\alpha_{\rm PPD}$, the efficiency of the viscous angular momentum transport outside the gap, is independent of the dust vertical diffusion.

\begin{figure}[H]	
	\begin{center}
		\includegraphics*[bb=0 0 270 290,scale=0.9]{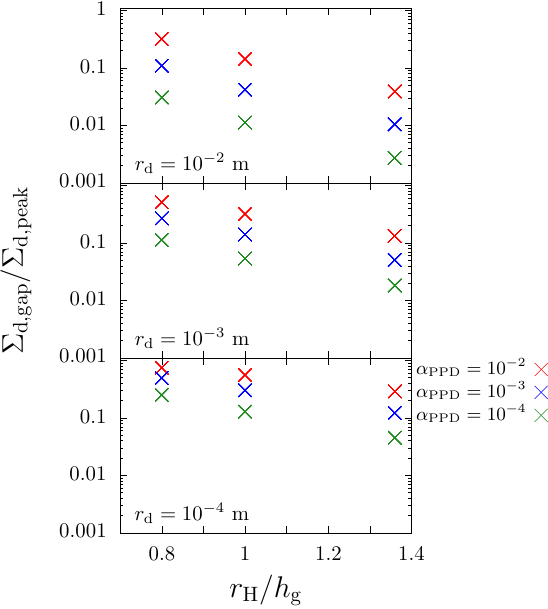}
	\end{center}
	
	\caption{Relationship between the depth of dust gap and $r_{\rm H}/h_{\rm g}$ obtained by the empirical formulae of \citet{z18}. Colors represent results for different values of the viscous parameter of the PPD $\alpha_{\rm PPD}$: red, blue, and green correspond to $\alpha_{\rm PPD}=10^{-2}$, $10^{-3}$, and $10^{-4}$, respectively. Top, middle, and bottom panels show the cases of $r_{\rm d}=10^{-2}$, $10^{-3}$, and $10^{-4}$~m, respectively.}
	\label{fig:Sdgap}
\end{figure}

\Figref{fig:Sdgap} shows the relationship between the depth of the dust gap and $r_{\rm H}/h_{\rm g}$. We find that the dust gap becomes deeper with increasing planetary mass and increasing dust size.

Now, we obtain the dust mass accretion rate onto the CPD from \Equref{eq:Macc}.
We assume that the dust surface density at the accretion band is equal to the value at the bottom of the dust gap. Then,
\begin{eqnarray}
\Sigma_{\rm d}=\Sigma_{\rm d, gap}=\frac{\Sigma_{\rm d, gap}}{\Sigma_{\rm d,peak}}f_{\rm peak}\Sigma_{\rm d,0}\\
f_{\rm peak} \equiv \frac{\Sigma_{\rm d, peak}}{\Sigma_{\rm d,0}},
\end{eqnarray}
where $\Sigma_{\rm d,0}$ is the dust surface density at the unperturbed region of the PPD.
Here, we assume $\Sigma_{\rm d,0}=\Sigma_{\rm d, MMSN}f_{\rm dust}$, where $\Sigma_{\rm d, MMSN}$ and $f_{\rm dust}$ are the dust surface density of the MMSN model at 5.2~au \citep{h81} and the depletion factor of the dust surface density ($f_{\rm dust}=1$ corresponds to the MMSN model), respectively. We further assume $f_{\rm dust}=1$ for simplicity.
The ratio between the peak and bottom values of the dust surface density $\Sigma_{\rm d, gap}/\Sigma_{\rm d,peak}$ is obtained using \Equref{eq:zhang18}.
Since the dust particles treated in this study are sufficiently small to avoid significant accumulation at the gas pressure bump, we assume $f_{\rm peak}=1$.

\begin{figure}[H]
\begin{tabular}{ccc}
\hspace{-2cm}
	\begin{minipage}{0.33\hsize}	
		\begin{center}
			\includegraphics*[bb=0 0 199 311,scale=0.9]{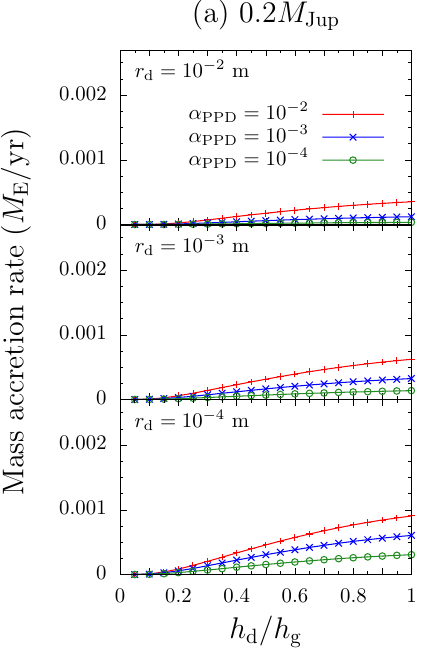}
		\end{center}
	\end{minipage}&
	
\hspace{0.2cm}
	\begin{minipage}{0.33\hsize}	
		\begin{center}
			\includegraphics*[bb=0 0 199 311,scale=0.9]{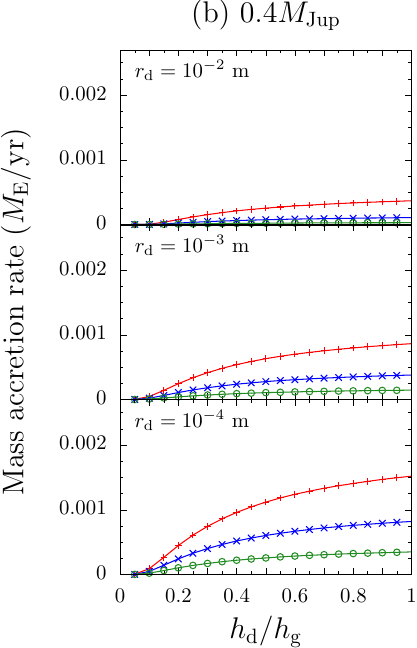}
		\end{center}
	\end{minipage}&
	
\hspace{0.2cm}
	\begin{minipage}{0.33\hsize}	
		\begin{center}
			\includegraphics*[bb=0 0 199 311,scale=0.9]{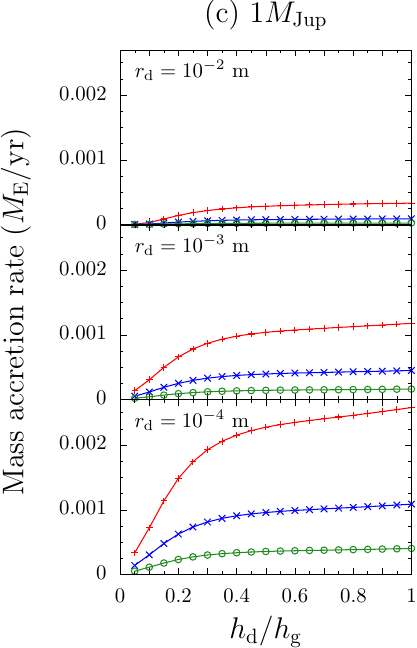}
		\end{center}
	\end{minipage}

\end{tabular}
	\caption{Dust mass accretion rate onto the CPD as a function of the dust scale height $h_{\rm d}$, where the effects of dust gap is taken into account. We assumed $f_{\rm dust}=1$ for simplicity, neglecting dust depletion in the PPD in the late stage of giant planet formation (see text for details). Colors represent the results for different values of the viscous parameter of the PPD $\alpha_{\rm PPD}$: red, blue, and green correspond to $\alpha_{\rm PPD}=10^{-2}$, $10^{-3}$, and $10^{-4}$, respectively. Top, middle, and bottom panels show the cases of $r_{\rm d}=10^{-2}$, $10^{-3}$, and $10^{-4}$~m, respectively.}
	\label{fig:Macc-hd}
\end{figure}

\Figref{fig:Macc-hd} shows the plots of the dust mass accretion rates as a function of $h_{\rm d}/h_{\rm g}$ for various values of planetary mass, particle size, and the viscosity parameter $\alpha_{\rm PPD}$ of the PPD.
It should be noted that we assume $f_{\rm dust}=1$ for simplicity in the calculation of the mass accretion rates presented here, the amount of dust in the PPD should be less than the MMSN value (i.e., $f_{\rm dust}\ll1$) in the late stage of planet formation. Therefore, it is unlikely that the core mass will become unnaturally large due to excessive dust supply, as would be obtained by simply multiplying the mass accretion rate shown here by the growth timescale of the planet.

We find that the mass accretion rate and its dependence on dust scale height are nearly independent of the planetary mass for $r_{\rm d}=10^{-2}$~m.
As we mentioned in \Secref{sec:Pcap-total}, $P_{\rm total,cap}$ in the case of strong vertical stirring of particles (i.e., $h_{\rm d}/h_{\rm g}\simeq 1$) becomes about three times larger when the planetary mass increases from $0.2M_{\rm Jup}$ to $1M_{\rm Jup}$ (\Figref{fig:Ptotal-hd}).
In this case, the planet's Hill radius increases by about a factor of 1.7, from $r_{\rm H}/h_{\rm g}=0.8$ to 1.36, thus $r_{\rm H}^2$ in \Equref{eq:Macc} increases by a factor of $(1.7)^2 \sim 3$. 
On the other hand, from the upper panel of \Figref{fig:Sdgap}, $\Sigma_{\rm d,gap}$ in the case of $r_{\rm d}=10^{-2}$~m becomes about one order of magnitude smaller when the planetary mass increases from $0.2M_{\rm Jup}$ ($r_{\rm H}/h_{\rm g}=0.8$) to $1M_{\rm Jup}$ ($r_{\rm H}/h_{\rm g}=1.36$).
As a result, the effects of increased planetary mass on the above three quantities that appear on the r.h.s. of \Equref{eq:Macc} nearly cancel out, resulting in the mass accretion rate that is nearly independent of the planetary mass in the case of $r_{\rm d}=10^{-2}$~m.
On the other hand, in the case of $r_{\rm d}=10^{-4}$~m, the change of $\Sigma_{\rm d,gap}$ due to the increase in the planetary mass is small, so the effects of increased $P_{\rm total,cap}$ and $r_{\rm H}$ with increased planetary mass are dominant. Therefore, the mass accretion rate significantly increases with increasing planetary mass in this case.
In all the cases shown in \Figref{fig:Macc-hd}, mass accretion rates are larger for smaller particles due to the shallower gap opening for smaller particles (i.e., $\Sigma_{\rm d,gap}$ becomes larger, as shown in \Figref{fig:Sdgap}).

\subsubsection{Ratio of Dust-to-Gas Inflow Rates}\label{sec:chi}
Using the results of this work together with the empirical formulae of the surface densities of gas and dust at the gap bottom \citep{k15,z18}, we obtain the ratio of dust-to-gas inflow rates onto the CPD. Using the total capture rate $P_{\mr{cap,total}}$ obtained in the present work and the gas mass accretion rate $\tilde{\dot{M}}_{\rm g}$ obtained by the hydrodynamic simulation of \citet{m22}, the ratio of dust-to-gas inflow rates is obtained by the following equation \citep{h20}: 
\begin{equation}
\chi = \frac{\dot{M}_{\mr{d}}}{\dot{M}_{\mr{g}}} = \frac{P_{\mr{cap,total}}\Sigma_{\mr{d}}h_{\mr{g}}}{\tilde{\dot{M}}'_{\mr{g}}\Sigma_{\mr{g,0}}r_{\mr{H}}},
\end{equation}
where we substituted the dust surface density at the dust gap obtained by \Equref{eq:zhang18_Sd} into $\Sigma_{\mr{d}}$.
In the above, $\tilde{\dot{M}}'_{\mr{g}}$ is the corrected gas surface density at the gap using the empirical formulae obtained by the global hydrodynamic simulation (Kanagawa et al. 2015; see also Maeda et al. 2022), which is obtained as 
\begin{equation}
\tilde{\dot{M}}'_{\mr{g}} = \tilde{\dot{M}}_{\mr{g}} \times \frac{\Sigma_{\rm g,glo}}{\Sigma_{\rm g,loc}},
\end{equation}
where $\Sigma_{\rm g,glo}$ and $\Sigma_{\rm g,loc}$ are the gas surface densities at the gap bottom obtained by the global simulation of \citet{k15} and the local simulation of \citet{m22}, respectively.

\Figref{fig:chi-hd} shows the ratio of dust-to-gas inflow rates as a function of $M_{\rm p}/M_{\rm Jup}$. In the case of small dust scale height ($h_{\rm d}/h_{\rm g}\lesssim 0.1$), the ratio of dust-to-gas inflow rates increases with increasing planetary mass. For example, in the case of $h_{\rm d}/h_{\rm g}=0.1$ and $\alpha_{\rm PPD}=10^{-2}$, the ratio of dust-to-gas inflow rates increases by about one order of magnitude when the planetary mass increases from $0.2M_{\rm Jup}$ ($r_{\rm}/h_{\rm g}=0.8$) to $1M_{\rm Jup}$ ($r_{\rm}/h_{\rm g}=1.36$). This reflects the fact that the dust accretion band extends down to lower altitudes for a large planetary mass (\Secref{sec:dd}), which significantly influences the dust accretion rates when the vertical stirring of dust particles is limited.
On the other hand, for the larger dust scale height ($h_{\rm d}/h_{\rm g}\gtrsim 0.3$), the ratio of dust-to-gas inflow rates is nearly independent of planetary mass or even decreases with increasing planetary mass. In this case, the effect of the deeper gap (thus lower dust surface density) for a larger planetary mass is dominant.
Thus, the planetary mass dependence of the ratio of dust-to-gas inflow rates into the CPD varies depending on the degree of vertical stirring of dust particles in the PPD.

\subsubsection{Implication for Satellite Formation Models}\label{sec:implication}
\begin{figure}[H]
\begin{tabular}{ccc}
\hspace{-2cm}
	\begin{minipage}{0.33\hsize}	
		\begin{center}
			\includegraphics*[bb=0 0 196 312,scale=0.9]{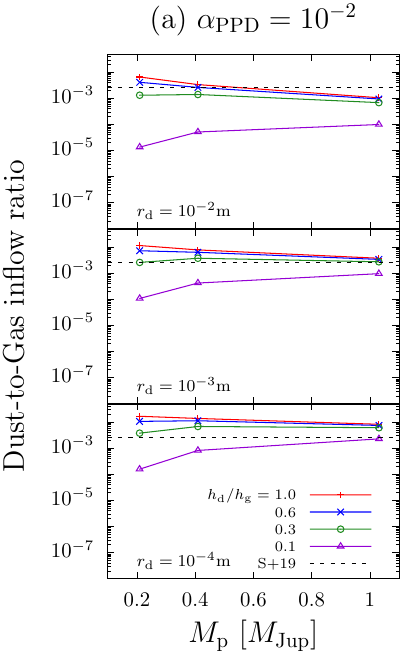}
		\end{center}
	\end{minipage}&
	
\hspace{0.2cm}
	\begin{minipage}{0.33\hsize}	
		\begin{center}
			\includegraphics*[bb=0 0 196 312,scale=0.9]{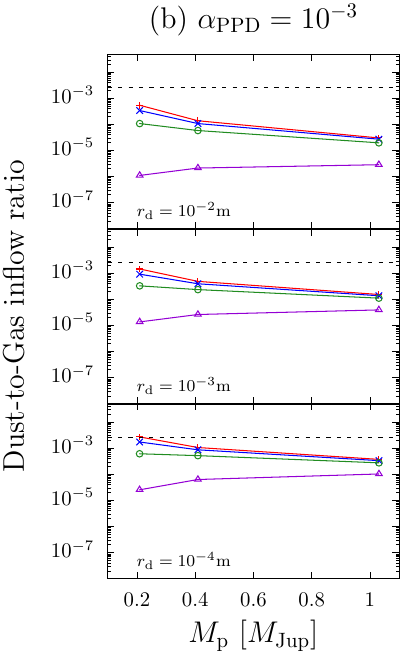}
		\end{center}
	\end{minipage}&
	
\hspace{0.2cm}
	\begin{minipage}{0.33\hsize}	
		\begin{center}
			\includegraphics*[bb=0 0 196 312,scale=0.9]{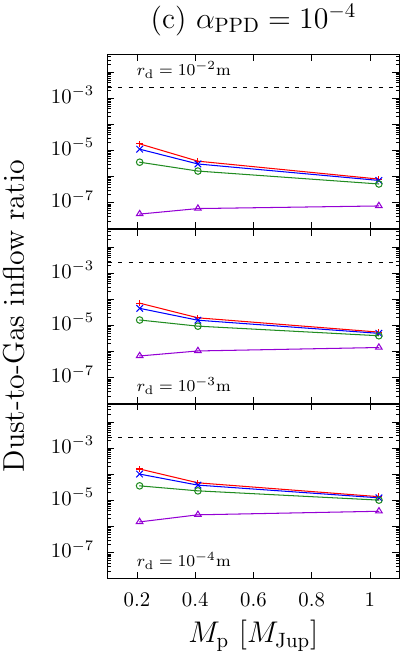}
		\end{center}
	\end{minipage}

\end{tabular}
	\caption{Ratio of dust-to-gas inflow rates onto the CPD as a function of $r_{\rm H}/h_{\rm g}$. Note that we assumed $f_{\rm dust}=1$ for simplicity, neglecting dust depletion in the PPD in the late stage of giant planet formation (see text for detail). Different colors represent different dust scale heights in the PPD: red, blue, green, and violet correspond to $h_{\rm d}/h_{\rm g}=$1.0, 0.6, 0.3, and 0.1, respectively. The dashed line corresponds to the value assumed in the Galilean satellite formation model of \citet{sh19}, $\chi=0.0026$.}
	\label{fig:chi-hd}
\end{figure}

\citet{sh17} investigated satellitesimal formation by calculating collisional growth of dust in a one-dimensional steady CPD. They found that satellitesimals can be formed if the ratio of dust-to-gas inflow rates is $\chi \simeq 1$. On the other hand, in the case of smaller $\chi$, they found that particles that have grown from small sizes with $St \ll 1$ to larger sizes with $St \simeq 1$ quickly fall to the central planet and cannot form satellitesimals.
\Figref{fig:chi-hd} shows that $\chi \lesssim 10^{-2}$ even when the PPD viscosity parameter is as high as $\alpha_{\rm PPD} = 10^{-2}$ and the dust scale height is sufficiently large, indicating that the conditions for the formation of satellitesimals in \citet{sh17} cannot be satisfied.
Recently, \citet{sm23} investigated satellitesimal formation in the CPD with magnetic wind-driven accretion and found that such CPDs can relax the condition for satellitesimal formation ($\chi \geq 0.02$).
However, even this relaxed condition seems difficult to satisfy in our results shown in \Figref{fig:chi-hd}.

On the other hand, \citet{sh19} proposed ``slow-pebble-accretion scenario'', which can reproduce major characteristics of the current Galilean satellite systems, such as their orbits, masses, and compositions.
In this scenario, several planetesimals captured by the CPD grow by accretion of pebbles that are continuously supplied to the CPD, while they slowly move inward in the CPD to form satellites. They assumed that the ratio of dust-to-gas inflow rates is $\chi=0.0026$, independent of the stages of the growth of the host planet.
We plotted the value of $\chi$ $(=0.0026)$ that corresponds to the one assumed in \citet{sh19} with the dashed lines in \Figref{fig:chi-hd}. \Figref{fig:chi-hd} shows that the value of $\chi$ assumed in \citet{sh19} can be achieved in cases where $\alpha_{\rm PPD}$ is large and/or the dust scale height is large, although the values of $\chi$ that we obtained vary somewhat depending on adopted dust size and planetary mass. It should be noted that the ratio of the dust-to-gas inflow rates would become lower if the depletion of solids in the PPD (i.e., $f_{\rm dust} < 1$; see Section 4.3.1) in the late stage of giant planet formation is taken into account.

According to the polarization observation of HD 163296 by ALMA, the dust scale height in the gap is estimated to be $h_{\rm d}=0.3-0.7 h_{\rm g}$ and the dust size is $r_{\rm d}\lesssim 10^{-4}$~m \citep{ok19}. From our results, the condition for satellitesimal formation obtained by \citet{sh19} can be achieved in cases where $\alpha_{\rm PPD} \sim 10^{-2}$ when $f_{\rm dust}=1$ is assumed.
\citet{c18} performed a population synthesis simulation of satellite formation around a Jupiter-mass planet using the results of radiative hydrodynamical simulation obtained by \citet{sz17}. In their model, dust particles inflow onto the CPD with the same profile as the gas with the ratio of dust-to-gas inflow rates ($0.03-0.5$), and the satellitesimals were assumed to grow via streaming instability. \Figref{fig:chi-hd} shows that such high ratio of dust-to-gas inflow rates ($\sim 10^{-2}-10^{-1}$) can only be achieved under limited conditions of $\alpha_{\rm PPD} \gtrsim 10^{-2}$, $h_{\rm d}/h_{\rm g}\simeq 1$, and $M_{\rm p}\simeq 0.2M_{\rm Jup}$ when $f_{\rm dust}=1$ is assumed.
However, it has recently been shown that planet-driven meridional circulation could increase the ratio of dust-to-gas inflow rates to $\sim 10^{-2}-10^{-1}$ \citep{sz22}. These conditions of high ratio of dust-to-gas inflow rates could be achieved by considering global flows in the PPD.

\citet{ba22} estimated the dust-to-gas mass ratio of the CPD candidate of AS 209b as $\lesssim 2\times10^{-4}-9\times 10^{-4}$, suggesting that the dust in the CPD is more depleted than the typical ISM environment. They inferred that the depletion is presumably due to limited supply and/or rapid radial drift of dust within the CPD. Our results support the former, although we cannot rule out the possibility of the latter.
It should be noted that the results presented here depend on the assumed dust/gas surface densities, which are described in \Secref{ap:mdot}.

In the present work, we assume the isothermal gas for simplisity and did not consider sublimation of icy dust or changes in dust composotion. For example, in the case of Jupiter's orbit, substituting $a=5.2$~au in \Equref{eq:T} yields 123~K. However, the temperature within the CPD may increase due to viscous heating \citep[e.g.,][]{cw02}, and effects of shock heating during accretion onto the CPD may also be important. In this case, if icy dust particle sublimate, the amount of solids supplied to the CPD could be smaller than that shown here. In our model, the composition of the dust only affects its assumed bulk density, and the results will not change significantly as long as the densities are similar. Further studies that take account of the changes in temperature and composition during accretion is important for better understanding of dust supply to CPDs \citep[e.g.,][]{rj20}.

\subsection{Capture Position of Dust Particles Compared to Planetesimals}
In \Secref{sec:cap-hist}, we examined the distribution of the capture positions of dust particles in the CPD (\Figref{fig:cap-hist}).
\citet{f13} examined capture of planetesimals by the CPD and obtained the critical radial distance from the planet for capture, i.e., those planetesimals passing within this critical radius are captured due to strong gas drag. \citet{so17} also examined orbital evolution of captured planetesimals in the CPD. These studies showed that meter- to kilometer-sized planetesimals can be captured in the region with $R \lesssim 10^{-2} r_{\rm H}$ for the gas-starved CPD, which roughly includes the current radial locations of the Galilean satellites and extends to more inner regions than the dust capture locations shown in \Figref{fig:cap-hist}. 
This is because planetesimals need sufficient energy dissipation to be captured by the CPD, which requires high gas density.
It should be noted that the capture radius as well as the maximum size of planetesimals that can be captured depend on the surface density of the CPD, and larger planetesimals can be captured in more outer regions if the CPD is more massive \citep{r18}.
Also, \citet{rj20} showed that most planetesimals that are captured by the CPD are strongly ablated due to frictional heating, resulting in the supply of dust grains to the CPD. They also showed that planetesimals smaller than 10~km would be ablated mainly at $R \gtrsim 10^{-2} r_{\rm H}$, providing dust particles in these regions.
These studies show that planetesimals are also important building blocks of satellites, but whether dust particles or planetesimals is the major contributor is outside the scope of this work.

\subsection{Model Limitations}\label{sec:caution}
In the present work, we investigated delivery of dust particles from the PPD to the CPD by performing orbital integration of dust particles under the influence of the gas flows in the vicinity of a planet obtained by high-resolution local hydrodynamic simulation, but some simplifications were introduced.
First, the local hydrodynamic simulation allows us to investigate the gas flow near the planet with high resolution, but as mentioned earlier, it is difficult to accurately represent the gap structures. We used the assumption of the surface densities of gas and dust based on the gap model derived from global simulation \citep{k15,z18} to estimate the mass accretion rates of gas \citep{m22} and dust (\Secref{ap:mdot}), and the ratio of dust-to-gas inflow rates (\Secref{sec:chi}).
However, direct use of global simulation is desirable to obtain realistic gap structures for both gas and dust simultaneously.
Second, in our simulation, the gas is assumed to be isothermal and inviscid for simplicity, and radiative transport is not included.
While we have focused on dust accretion into the CPD from a dynamical point of view assuming the locally isothermal PPD and CPD (\Equref{eq:T}) in this work, the temperature distribution in the CPD and its evolution is important for studies of compositional evolution of dust particles and forming satellites in the CPD \citep[e.g.,][]{dp15,rj20}.
Third, despite the assumption that the gas flow is laminar in the unperturbed state, we assume that dust particles are vertically stirred by some mechanism in the PPD. 
We treated the degree of vertical stirring as a parameter using the dust scale height, which is assumed to be independent of the efficiency of the angular momentum transport of the PPD.
Furthermore, although the influence of the gas flow on the orbital motion of dust particles is taken into account, the back reaction that dust exerts on the gas is not considered.

Recently, treating these effects more realistically, global simulations of PPDs with gas-dust two-component fluids have been performed \citep[e.g.,][]{bin21,sz22}. The mechanisms that can stir up dust particles have also been investigated in detail, such as MRI \citep{bh91} or meridional circulation \citep{fc16}. 
\citet{sz22} performed three-dimentioanl global hydrodynamic simulation that took into account radiative transport, gas viscosity, and back reactions from dust to gas, and investigated the dust supply to the CPD.
Their results showed that vertical stirring of dust particles by spiral wakes excited by the planet's gravity can play an important role in supplying solids to the CPD.
\citet{c23} performed three-dimentioanl global radiative magnetohydrodynamics (MHD) simulations. They found that a meridional circulation exists in the radiative MHD case, but high turbulent viscosity prevents gap opening, leading to a higher accretion rate than the hydrodynamic case. 
These simulations require a large amount of computer resource, but simulations with sufficiently high resolution will be important to examine, for example, distribution of captured solids in the CPD.
It will also be important to clarify the degree of dust retention and its dependence on various parameters using global simulation with sufficiently high resolution.
\citet{l23} performed three-dimensioanl global simulations of PPDs over gap-opening timescales and explore the planetary mass dependence of mass accretion rates onto the planet. They found that the scaling law is different from the local simulation of \citet{m22}, probably due to the shearing box approximation not considering the global accretion of the PPD. This effect may also affect dust accretion.

Our calculations do not take into account the size distribution of dust particles. Recently, \citet{k23} performed multiple dust grain size simulations and showed that the dust behavior itself does not change with or without consideration of size distribution. This indicates that the superposition of the results of single-size simulations can be used to obtain the dust accretion rate onto the CPD in the presence of dust size distribution.

%% file: 05conclusion.tex
In this work, we performed orbital integration of solid particles approaching a planet, taking account of drag from the gas flow perturbed by the planet obtained by \citet{m22} using three-dimensional hydrodynamical simulations of a local region around a planet.
We investigated the planetary-mass dependence of the orbits, capture positions, and capture rates of dust particles accreting onto the CPD. In addition, we examined the planetary-mass dependence of the degree of dust retention in accreting gas onto the CPD, which determines the ratio of dust-to-gas inflow rates, a key parameter in models of satellite formation in the CPD.
The main results are as follows:
\begin{enumerate}
\item{Similarly to the gas accretion band examined in our previous work \citep{m22}, the initial orbits in the PPD of particles captured in the CPD distribute above the midplane continuously in the $z$-direction in a zonal pattern except near the midplane. For dust particles with $St \ll 1$, the radial width of the dust accretion band measured in units of the planet's Hill radius becomes wider with increasing planetary mass.}
\item{Dust particles accreting with the gas are supplied into a radially extended region of the surface of the CPD. Smaller particles tend to be captured in more inner regions of the CPD. Also, dust particles are captured in radially wider regions of the CPD in the case of a larger planetary mass.}
\item{The degree of dust retention in accreting gas onto the CPD increases with increasing planetary mass for a given dust scale height in the PPD. For the case of a small planet ($M_{\rm p}=0.2M_{\rm Jup}$), particles with insufficient altitude in the PPD cannot get assisted by the accreting gas flow and cannot accrete onto the CPD. On the other hand, for the case of a massive planet ($M_{\rm p}=1M_{\rm Jup}$), even in the case with a small dust scale height of $h_{\rm d}=0.1h_{\rm g}$, the ratio of dust-to-gas inflow rates in the case of particles with $r_{\rm d}=10^{-4}$ to $10^{-2}$~m can be as large as $10-30$\% of the dust-to-gas mass ratio in the PPD (\Figref{fig:dd-mp}).}
\end{enumerate}

The results of this study can be used to develop models of satellite formation around gas giants of various masses, e.g., Saturn, Jupiter, and exoplanets.
Further investigation on the delivery of dust particles onto the CPD using improved numerical models, such as global simulation of dust and gas components taking into account of their back reactions will be desirable for better understanding of dust supply onto the CPD (\Secref{sec:caution}).
Also, the amount of dust supplied to the CPD is highly dependent on the dust scale height in the PPD, the structures of the gas and dust gaps, and the size distribution of dust particles in the PPD. 
Further observational studies on these quantities as well as their planetary mass dependence will be desirable to derive more information.
In particular, given that giant planets are expected to cause meridional circulation \citep{t19,bi21,sz22}, it is highly important to observe PPDs in the late stages of giant planet formation.
Recently, CPD candidates around exoplanets have been detected, such as PDS70c and AS 209b \citep{i19,ba22}, and a comparison of model calculations of dust evolution in the CPD with observations of dust thermal emissions is also being attempted (Shibaike \& Mordasini, submitted).
These theoretical and observational studies will contribute to the development of more realistic satellite formation models.